# Electromagnetic Environment Analysis of High-Power Wireless Charging Device

ZHANG Zhengyang[1], LIU Zhihui[2*], ZHANG Wenjin[3], ZHANG Rui[4], XIAO Xiang[5]

1. Beihang University, Beijing 100083 China
2. Beijing Radiation Safety Technology Center, Beijing 100089 China * Corresponding author.
3. The High School Affiliated to Beijing Institute of Technology, Beijing 100086 China
4. University of British Columbia, Vancouver, V6T 1Z41.
5. Beijing Animal Husbandry General Station, Beijing 100107, China.

**Abstract**

**Objective** Aiming at the problems of many interference factors in the electromagnetic radiation simulation of electric vehicles, a field measurement scheme for charging devices is designed. Through the monitoring of wireless charging equipment, the radiation level and distribution of the electric field value and magnetic field value around the charging equipment is explored, to analyze the influence law of the electromagnetic environment. **Method** This paper introduces the principle and development status of electric vehicle charging, and analyzes the classification and methods of wireless charging. In this paper, the electric field and magnetic field of cars and minibuses are monitored, at positions such as: around the body, attenuation section, inside the car and inside the minibus. **Result** The range of electric field strength is 0.9 V/m to 48.1 V/m for cars, and 0.8 V/m to 74.7 V/m for minibuses. The electric field strength decays rapidly with the increase of the distance from the vehicle body, and the law is obvious. The range of magnetic induction intensity is 0.12μT-12.70μT for cars, and 0.15μT-27.06μT for minibuses. The magnetic induction intensity decays rapidly with the increase of the distance from the vehicle body, and the law is obvious. **Conclusion** This paper explores the radiation level and distribution of the electromagnetic field around this type of charging equipment. It is recommended that manufacturers of wireless charging devices for electric vehicles strengthen research on electromagnetic radiation shielding and take corresponding measures to control the level of electromagnetic radiation in areas accessible to the public.

## 1 Introduction

Due to the shortage of energy and the aggravation of environmental pollution, it is necessary to vigorously develop new energy electric vehicles. Therefore, the state has given a series of preferential policies. However, due to limited battery capacity, insufficient charging stations, and insufficient battery life, the public's acceptance of electric vehicles is insufficient. At present, the charging related issues of electric vehicles is one of the main bottlenecks restricting the development of new energy electric vehicles.

Due to the support of national industrial policies, charging stations have sprung up one after another, and then technical problems that need to be solved in wireless charging technology have arisen as the times require. Wireless charging technology [1] refers to a new technology that transmits energy through electric fields, magnetic fields, etc. without using wires. At present, the energy transmission technology of wireless charging technology can be divided into 6 types. The main consideration is to use the inductive power transfer system, based on its good functionality, high reliability, safe use, long life, plus non-contact and non-wear characteristics [2]. This paper studies the electromagnetic induction wireless charging technology.

With the promotion and application of wireless charging technology in electric vehicles, the electromagnetic environment safety issue has also attracted widespread attention. Electromagnetic compatibility is one of the directions: in-vehicle equipment in an electromagnetic environment will suffer from electromagnetic interference and work abnormally [3]. Geomagnetic detection technology has just shown its edge, and it continues to develop along with the huge market demand. The transmission power of electric vehicle wireless charging is very large, therefore, it needs to be evaluated by electromagnetic radiation dosimetry [4].

Health hazards are another direction: the health of the human body will be affected in the electromagnetic environment [5-7], such as long-term exposure, whether there will be blood changes [8].

Many people of insight in the industry have done a lot of research [9-26], and are also thinking about corresponding exploration methods. Such as: a safety application device based on a geomagnetic sensor CN201974905U.

## 2. Electromagnetic environment monitoring

2.1 Monitoring method:
Refer to the standards of GB8702-2014 and GB/T18387-2017 for measurement.

2.2 Monitoring objects:

Table 1 Basic situation of wireless charging monitoring

| Model | Applied Vehicles | Charging Frequency | Charging Power | Measurement location | Usage scenario |
|---|---|---|---|---|---|
| ZXWPT006 | car | 78kHz | 6kW | chassis rear | underground parking garage |
| ZXWPT030 | Minibus | 42kHz | 30kW | under the middle of the chassis | open air parking lot |

We have considered different power, different models, and different usage scenarios, see Table 1:

2.3 Monitoring layout

2.3.1 Three working conditions of point layout:

(1) When the vehicle and wireless charging device are completely powered off, the background electric field value and the background magnetic field value are monitored.

(2) When the wireless charging device is powered on but not working, the electric field value and magnetic field value are monitored.

(3) When the wireless charging device is working and adjusted to the maximum output power, the electric field value and magnetic field value are monitored.

2.3.2 Monitoring methods, monitoring points and distances:

(1) Firstly, the frequency range of the monitoring object is determined through frequency scanning.

(2) Monitoring at a distance of 20cm around the vehicle body. The main monitoring points: the front of the vehicle, the rear of the vehicle, and near the door, respectively at different heights of 0.5m, 1m, and 1.5m.

(3) To find the maximum monitoring value (the most unfavorable position accessible to the public), then to conduct section monitoring at the maximum value.

(4) In-vehicle monitoring: monitoring the driving position and the maximum value in the vehicle.

2.4 Monitoring layout

See Figure 2 and Figure 3 for details on the location of monitoring points:

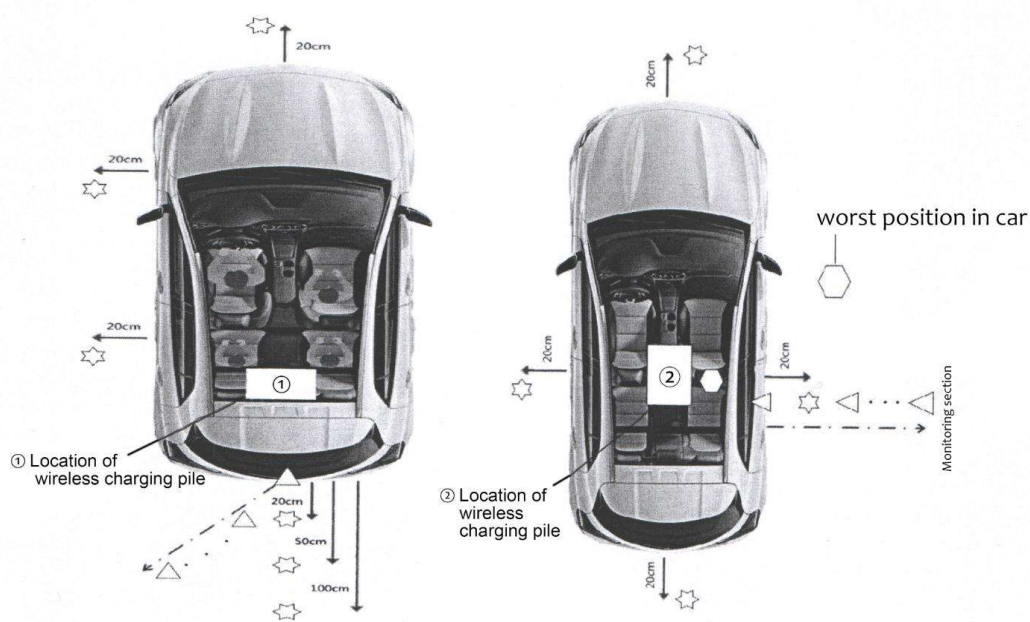

Figure 2    Schematic diagram of car monitoring layout

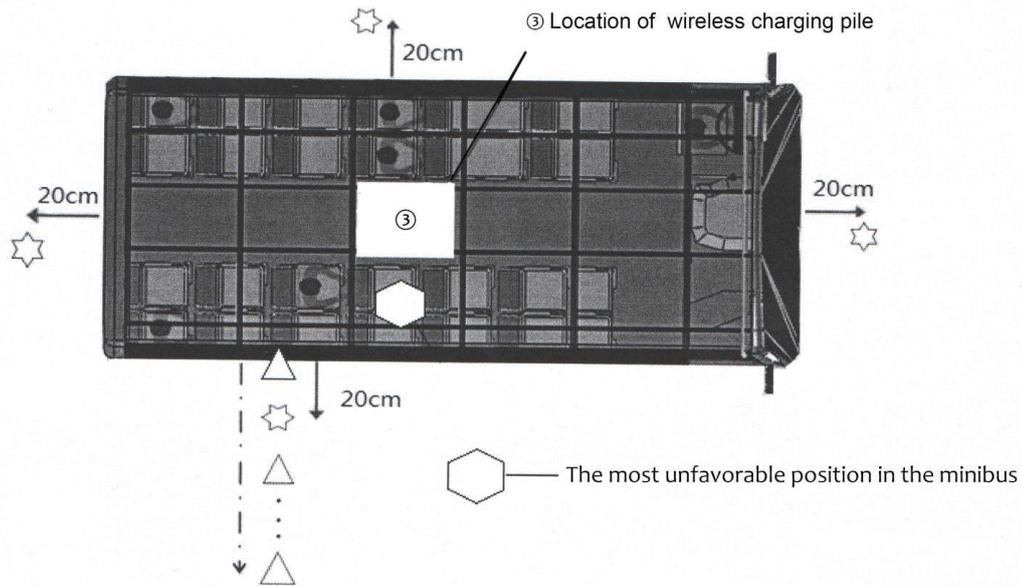

Figure 3　Schematic diagram of monitoring points for minibuses

## 3.Results and data analysis of electromagnetic radiation environmental monitoring

3.1 Monitoring results
3.1.1 Standby monitoring results

Table 2　Monitored results in standby state

| Position | Electric field value (V/m) | Magnetic field value (μT) | Remarks |
|---|---|---|---|
| directly behind the car | 0.74 | 0.12 | |
| background value | 0.71 | 0.11 | |
| GB 8702 limit value | 51.28 | 0.15 | |

3.1.2 Monitoring results of ZXWPT006 device (applied to cars)

　　GB 8702 limit value: the electric field value is 51.28 V/m, and the magnetic field value is 0.15μT. The background electric field value of the car is 0.71 V/m, and the background magnetic field value is 0.11 μT..The car monitoring results are shown in Figure 4 and Table 3:

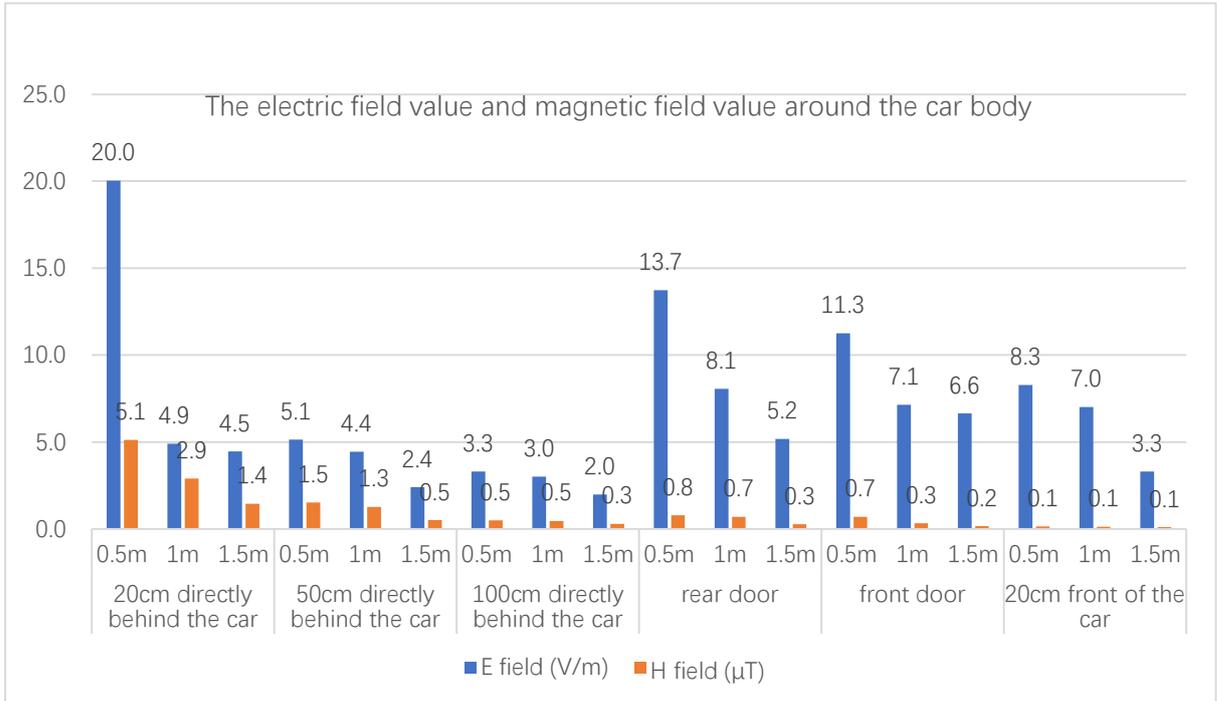

Figure 4-1　Monitoring results around the car body

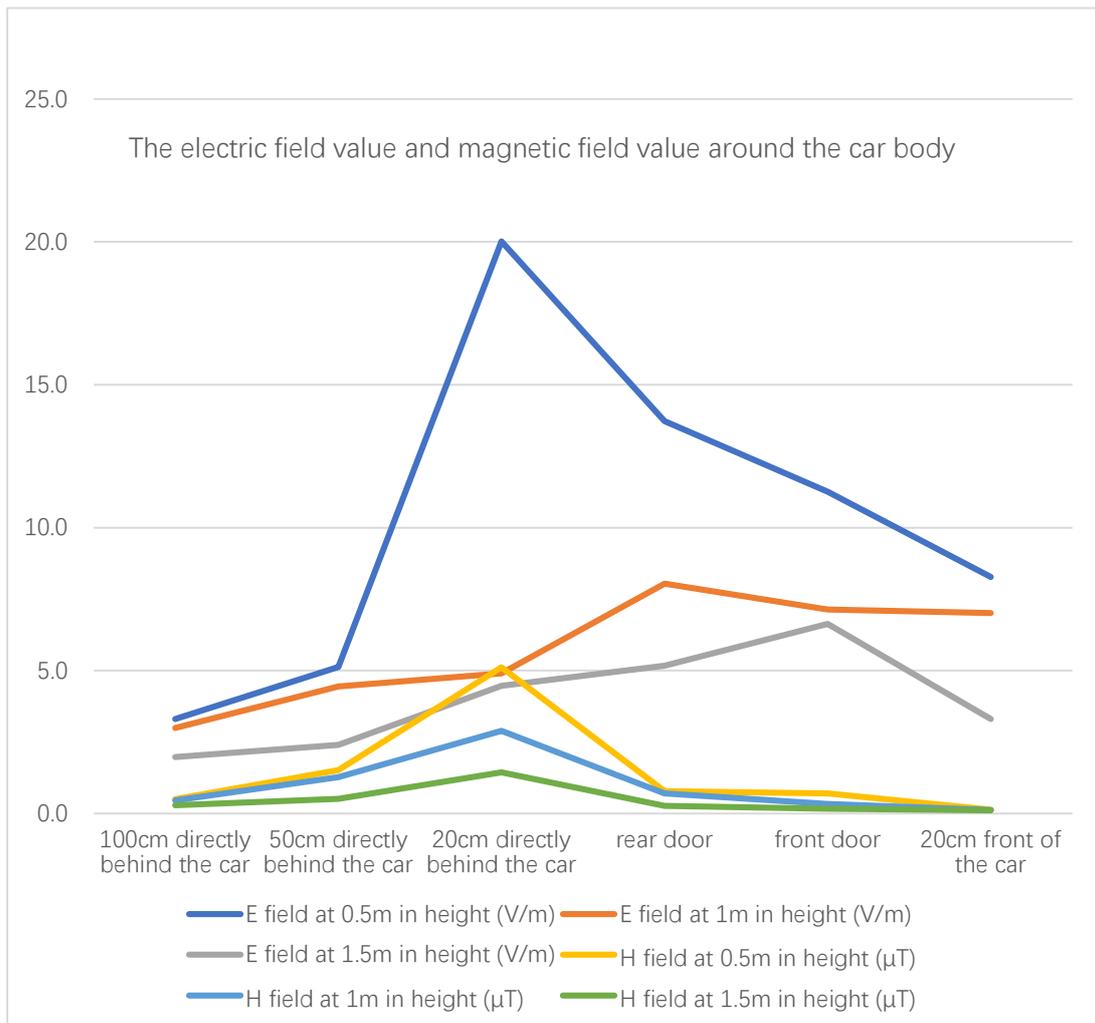

Figure 4-2　Monitoring results around the car body

When the height is 0.5 meters at 20cm behind the car, the electric field value is 20.1V/m.Which is relatively high, but it does not exceed the limit value of 51.28 V/m in GB 8702. The electric field values at other points do not exceed the standard. The magnetic field value is 5.2μT, exceeding the limit value of GB 8702 by 0.15μT. The magnetic field value exceeds the standard at heights of 0.5 meters, 1.0 meters, and 1.5 meters. There is electromagnetic influence at the distance behind the vehicle within 250cm. It falls within the standard limit only outside the range of 250cm around.

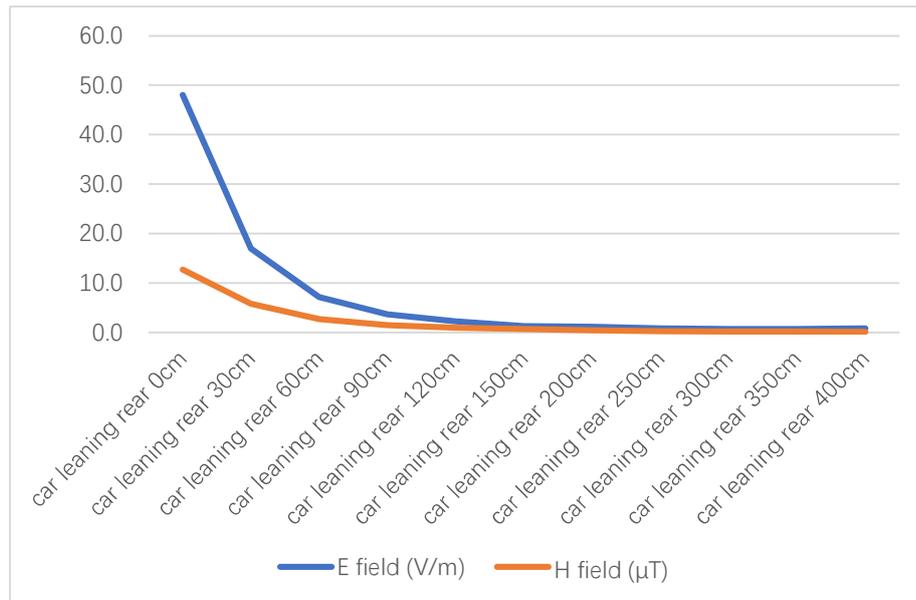

Fig.5 Electric field value and magnetic field value on the attenuation surface of the car body

Measured on a plane with a height of 0.15 meters around the car, the electric field value is 48.1 V/m, which is relatively high and close to the limit value of 51.28 V/m in GB 8702. The electric field values of other points also did not exceed the standard. The magnetic field value is 12.7μT, exceeding the limit value of GB 8702 by 0.15μT. The magnetic field value has electromagnetic influence within 250cm around the rear of the car, and it is lower than the limit value outside 250cm.

Table 3　Monitored results in the car

| No. | Position | Height | Electric field value (V/m) | Magnetic field value (μT) |
|---|---|---|---|---|
| 1 | driver's seat | heart position | 0.78 | 0.12 |
| 2 | Passenger on the left side of the car | heart position | 0.71 | 0.11 |
| 3 | middle rear | Chassis 0.15 m | 0.75 | 0.13 |

At driver's heart position, the electric field value is 0.78 V/m, and magnetic field value is 0.12μT. At passenger' heart position on the left side of the car, the electric field value is 0.71 V/m, and magnetic field value is 0. 11μT.At middle rear on 0.15m high, the electric field value is 0.75V/m, and magnetic field value is 0.13μT.

3.1.3 Monitoring results of ZXWPT030 device (applied to minibus)

The background value of the electric field value of a minibus is 0.80 V/m, and the background value of the magnetic field value is 0.15 μT; GB 8702 limit value: the electric field value is 70 V/m, and the magnetic field value is 0.29 μT. The monitoring results of minibuses are shown in Figure 6-Figure 7 and Table 4.

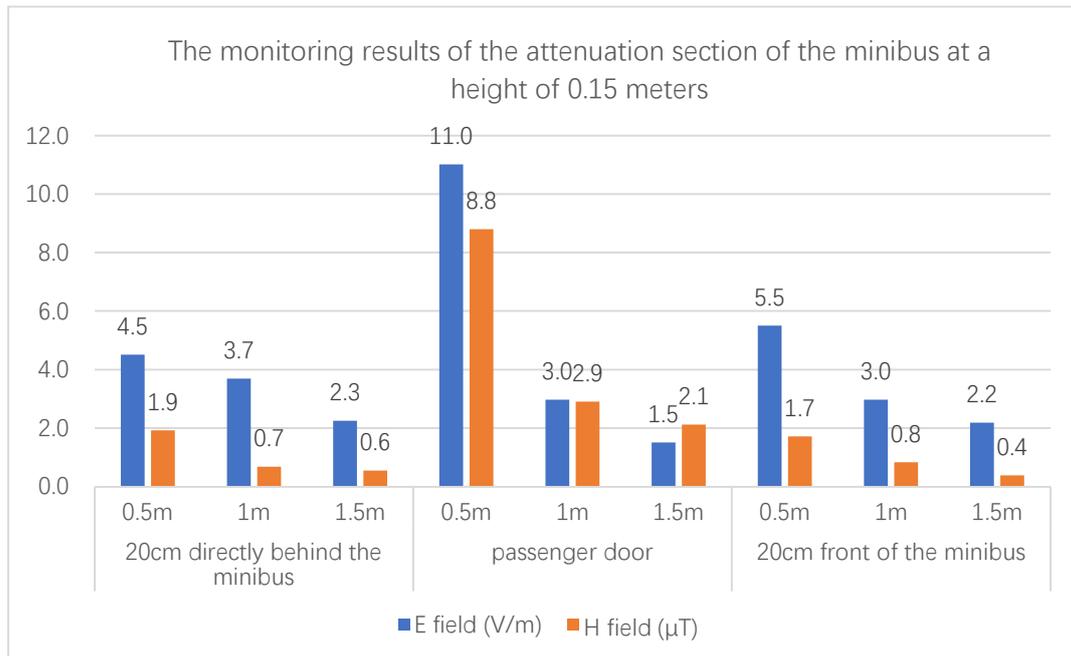

Figure 6 Electric field and magnetic field values at 0.5, 1.0, and 1.5 meters around the minibus

When measuring the height of 0.5 meters at a distance of 20cm from the side of the minibus (passenger boarding space), the electric field value is 11.0V/m, which is relatively high, but lower than the electric field limit value of 51.28 V/m in GB 8702, and other electric field values are also not Exceeded the standard. The magnetic field value is 8.8μT, exceeding the limit of GB 8702 magnetic field value by 0.15μT. The magnetic field value is at a height of 0.5 meters, 1.0 meters, and 1.5 meters, and the 20cm behind the car and 20cm in front of the car all exceed the standard limit of the magnetic field value. There is an electromagnetic radiation effect on the surrounding 20cm range.

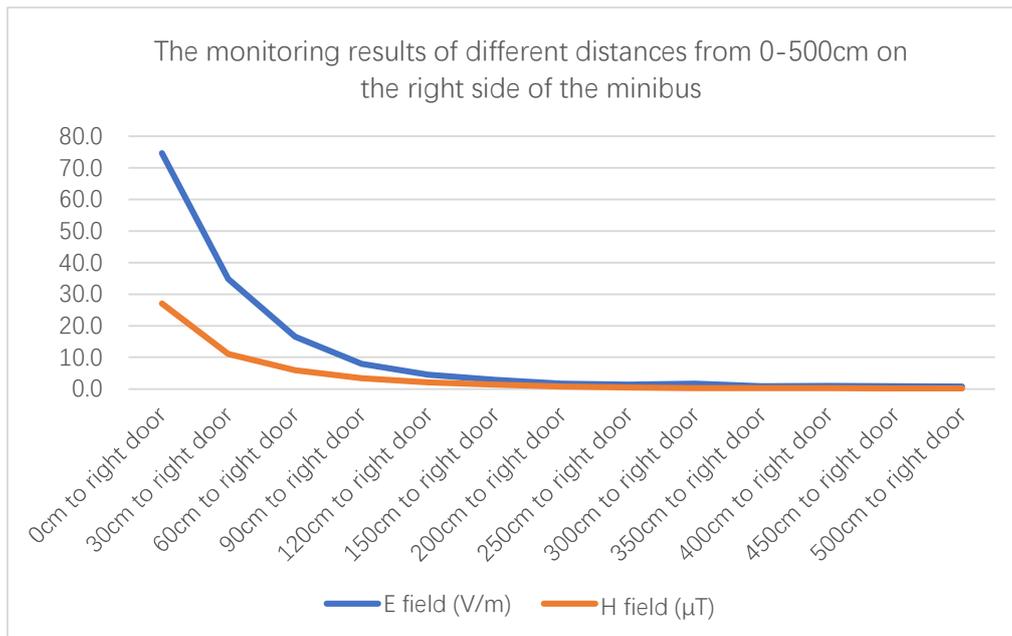

Fig. 7 Electric field value and magnetic field value of 0.15m attenuation section of minibus

When the height of 0.15 meters is measured at 0cm in the center of the side of the car (passenger boarding space), the electric field value is 74.8V/m, which is relatively high, exceeding the limit value of 51.28 V/m in GB 8702, and other electric field values do not exceed the standard. The magnetic field value is 27.1μT, exceeding the limit value of GB 8702 by 0.15μT.

Table 4　Monitored results in minibuses

| No. | Position | Height | Electric field value (V/m) | Magnetic field value (μT) |
| --- | --- | --- | --- | --- |
| 1 | In the center of the minibuses (above the charging station) | Chassis 0.15 m | 1.02 | **1.42** |
| 2 | driving seat | heart position | 0.91 | **0.33** |
|  | background value |  | 0.80 | 0.15 |
|  | GB 8702 limit value |  | 70 | 0.29 |

Above the charging station high 0.15m, the electric field value is 1.02V/m, and magnetic field value is 1.42μT. At driving' heart position, the electric field value is 0.91V/m, and magnetic field value is 0. 33μT.The magnetic field value is measured at a height of 0.15 meters, and has electromagnetic radiation effects on the surrounding range of 400cm.

# 4. Results

When the wireless charging device is only powered on and not working, the electromagnetic radiation environmental monitoring results tend to the background value.

4.1 Analysis of the electric field strength of the surrounding environment of the car during charging.

The electric field intensity monitoring value of the surrounding environment of the car (low-power wireless charging device) ranges from 0.9 V/m to 48.0 V/m, and the electric field intensity of 51.28 V/m in the electromagnetic environment control limit (GB8702-2014) is lower than 78kHz. limit.

The electric field intensity monitoring value of the surrounding environment of minibuses (high-power wireless charging devices) ranges from 0.8 V/m to 74.7 V/m. Except for the maximum value slightly exceeding the limit value, the electromagnetic environment control at other points is lower than 42kHz The limit value of the electric field strength of 70V/m in the limit value (GB8702-2014). The maximum value is located at 0cm in the center of the side of the minibus and at a height of 15cm.

The electric field strength attenuates rapidly with the distance from the car body, and the attenuation law is obvious. At 4m obliquely behind the car, the electric field intensity is close to the background level at a distance. At a distance of 5m from the side center of the minibus, the electric field intensity is close to the background level.

4.2 Analysis of the magnetic induction intensity of the surrounding environment of the car during charging

The monitoring value of the magnetic induction intensity of the surrounding environment of the car (low-power wireless charging device) ranges from 0.1μT to 12.7μT, exceeding the limit value of the magnetic induction intensity of 0.15μT in the electromagnetic environment control limit at 78kHz (GB8702-2014).

The monitoring value of the magnetic induction intensity of the surrounding environment of the minibus (high-power wireless charging device) ranges from 0.15μT to 27.06μT, and some of them exceed the limit value of the magnetic induction intensity of 0.29μT in the electromagnetic environment control limit at 42kHz (GB8702-2014).

The magnetic induction intensity decays rapidly with the increase of the distance from the vehicle body, and the decay law is obvious. The magnetic induction intensity drops below the limit at a distance of 3.5m from the oblique rear of the car, and is close to the background level at 4m; background level.

4.3 Environmental Impact Analysis of Electromagnetic Radiation in Vehicles

The electric field intensity monitoring value at the heart position of the driver's seat is 0.8 V/m, and the magnetic induction intensity monitoring value is 0.1 μT; the magnetic induction intensity monitoring value at the heart position on the rear left side of the car is 0.11 μT, and the middle distance of the rear row inside the car is 0.15 m from the chassis The monitoring value of the magnetic induction intensity at the place (the place closest to the wireless charging device in the car) is 0.13μT. For small cars (low-power wireless charging device), due to the wireless charging device is located at the rear of the car, and the charging power is low, the car body shielding is good and other factors, the monitoring values of the electric field intensity and magnetic induction intensity in the car are close to the background value, lower than the limit value of GB8702 .

The monitoring values of the electric field strength inside the minibus are all lower than the limit value. The monitoring value of the magnetic induction intensity is relatively large, and the maximum value is located on the ground in the car directly above the charging device, which is 1.4μT, which is higher than the limit value of GB8702.

## 5. Discussion and conclusion

The research on the electromagnetic radiation intensity distribution of electric vehicle charging pile sites is an exploration of cutting-edge technology and the basis for the research on personal electromagnetic safety in the electric vehicle environment. According to the requirements of relevant domestic and foreign standards, and IEEE standards on human body exposure to electric field, electromagnetic and electromagnetic field safety level standard draft [27], a real vehicle measurement scheme was designed and measurement analysis was carried out. It is concluded that the electric field intensity in the electric vehicle decreases with the increase of the horizontal height, and the higher the height, the faster the decrease. The electric field strength inside the electric vehicle decreases with the increase of the horizontal distance, and the farther the distance is, the faster the decrease is.

High-power wireless charging devices have greater electromagnetic radiation environmental impact than low-power wireless charging devices. Regardless of whether it is inside or outside the car; due to airtightness, the electromagnetic environment impact of wireless charging for cars with a high chassis is greater than that for cars with a bottom chassis.

In the public accessible area, the closer the body is to the wireless charging device, the smaller the obstruction to the wireless charging device, and the greater the impact of electromagnetic radiation on the environment. The maximum value appears at the position closest to the wireless charging device, and the height is between the ground and the chassis of the vehicle.

Geomagnetic monitoring is affected by many factors, such as water leakage, obstruction of obstacles, communication signal interference, etc. This measurement does not have these effects. This measurement is equivalent to an outdoor simulation. Therefore, the actual working conditions will be much more complicated. How to consider more relevant factors when building charging facilities is worth further discussion.

With the intensification of the energy crisis and the strengthening of environmental protection awareness, the popularity of pure electric vehicles will become wider and wider. Therefore, manufacturers and research institutions should invest manpower and material resources in relevant research to effectively improve the level of electromagnetic compatibility and electromagnetic interference suppression. When building a charging site, the factors that should be considered are how to avoid artificial interference sources with large electromagnetic interference, how to prevent water leakage caused by rain, and how to screen different car body materials [28] to reduce electromagnetic radiation influences.